# Black Hole in Binary System is the Source of Cosmic Gamma-Ray Bursts and Their Counterparts

**Alexander V. Kuznetsov**
Space Research Institute, Russian Academy of Sciences, Moscow, Russia
akuznetsov@rinet.ru

**Abstract**

Discovery of GRB clusters allows us to determine coordinates and characteristics of their sources. The objects radiating GRBs are reliably identified with black hole binaries, including the Galactic binaries. One of the unusual GRB properties, which are determined by black hole, is revealed in that the measured arrival direction of GRB does not coincide with the real location of its source. Just this fact allows us to find the objects radiating GRBs. On the basis of the general relativity theory's effects, observed in the GRB clusters, the technique of the black hole masses measurement is developed. The calculated black hole masses for the majority of known Galactic BH binaries are presented. It is briefly shown how the incorrect interpretations of observational facts result in an erroneous idea of the GRB cosmological origin. In fact, two problems are solved in the paper: the GRB origin and the reality of BH existence.

**1. Introduction**

Intensive study of cosmic gamma-ray bursts (GRBs) have been carried out since 1973. They were observed for the first time as pulse gamma radiation in the energy range of 0.3 - 2 MeV with the burst duration of ~ 10 s [1].

In the initial period of study it was natural to suppose that the bursts had Galactic origin, but this supposition was not experimentally confirmed. The large-scale anisotropy of GRBs, which was expected in the case of Galactic origin, was not really observed. The distribution of GRBs was found to be high isotropic [2] that indicated their possible extra-Galactic origin.

The GRB counterparts were detected for GRB 970228 in the optical [3] and X-ray [4] energy ranges. For GRB 970508 the glow was observed in the radio and optical ranges [5]; from the spectrum of optical glow the redshift of $0.835 \lesssim z \lesssim 2.3$ was measured [6]. For the optical counterpart of GRB 970508 were determined the coordinates $\alpha_{2000} = 06^h\,53^m\,49^s.43$, $\delta_{2000} = +79° \, 16'\, 19.6"$ with uncertainties of ~ 1 arcsec [7]. The Hubble Space Telescope observations reveal that optical glow is radiated from a pointlike source [8]. For today, the $z$ values measured on the optical glow for several tens GRBs ranges from 0.44 (GRB 060512) to 6.7 (GRB 080913). The redshifts for the events associated with supernovae (SN), as well as those determined indirectly, from the host galaxy, in that number, are not taken into account here.

The measurement accuracy of GRB arrival directions from the optical and radio glow reaches of angular second. However, this accuracy ensures at best identification with a "host" galaxy up to the lowest possible stellar magnitude of ~ 28 mag. In many cases the GRBs with known arrival direction do not find its "host" galaxy. This observational fact is known, as the "no host galaxy" problem [9].

The measurements of arrival directions for several thousand GRBs have shown the absence of events having identical coordinates [10, 11]. A simple interpretation of this observational fact is that the repeated GRBs do not exist and, accordingly, the sources of the repeated GRB do not exist as well. This supposition results in idea of GRB generation in a powerful explosive process, which destroys the radiation source. These suppositions have not been confirmed neither theoretically nor experimentally but, as it will be seen, they founded a base for erroneous concept about the GRB origin. At the same time, in 1995-1997 the papers were appeared suggesting the existence of clusters of repeated GRBs with unusual properties [12, 13]. One of these properties is that the observed repeated GRBs do not coincide in position with their source, as well as with each other.

The discovered GRB clusters in a number about 100 were studied and the locations of their sources have roughly been determined [14]. This list, which also contains some characteristics of clusters, can be considered as an attempt to catalogues of GRB sources. It will be shown below that all existing catalogues are the GRB catalogues, which carry no information about the location of GRB sources virtually. Search of the clusters has been performed on basis of the GRB catalogue [15] of the Burst and Transient Source

Experiment (BATSE) of the Compton Gamma-Ray of GRB arrival directions in the experiments BeppoSAX (www.asdc.asi.it/bepposax/), All-Sky Monitor (ASM) on the RXTE satellite [16], and the interplanetary spacecraft network - IPN (ssl.berkeley.edu/ipn3/index.html) were also used.

The basic results [14] are briefly reviewed below. The GRB clusters are singled out statistically in time and space. The physical relationship between the GRBs in clusters shows that they are emitted by a same source. The characteristic size of clusters is about of 30 angular degrees, the number of events is ~ 7, on the average, and the time, during which these events are observed, equals 3 - 4 months. The examples of clusters are given in Figs. 1, 3, 4, and 5. The cluster maximum size determined by the experienced way equals 36°. Gamma-ray bursts in clusters are located, as a rule, in the ring-shaped area around the source, which is in the center of a cluster. The size of cluster and the error boxes of GRB locations define that the events in cluster do not coincide on location with the source, as well as with each other. The indicated feature of the GRB clusters is, evidently, the basic reason why the repeated GRBs were not discovered. Strictly speaking, the direction of GRB arrival is determined at the measurement point and this GRB arrival direction is automatically (as usual) identified with the source location. However, for the GRB phenomenon the identification of the GRB arrival direction with the location of its source is an artifact. In the existing identification of GRB arrival direction with the location of its source, their coordinates can differ up to 18 angular degrees. It is evident that for the correct localization of GRB sources must be used the GRB clusters observed from these sources.

The lack in the universe of astronomical objects, which have the angular size up to 36 degrees, allows us to conclude that the angular size of GRB cluster does not determine the real size of a source. The cluster should be considered as the gravitational image of the GRB source.

Using the coordinates of GRB sources defined by cluster's center to the accuracy of about 5 degrees [14] the GRB sources have been sought according to the list of candidates to black holes [17], as well as by the catalogue of gamma- ray sources [18]. Search of the sources resulted in identification of clusters with the Galactic black hole binaries (BHBs) GRO J0422+32 and GS 1354 − 645, as well as with the nearest-neighbor galaxies M31, M33, M81.

The observed properties of GRB clusters result in the following conclusions: 1) The GRB clusters represent actually the clusters of repeated GRBs and, therefore they allow us to determine the position of their source; 2) The measured arrival direction of GRB and its counterpart does not coincide with the position of their source. The GRB arrival direction represents an apparent position of a source. So, the true host galaxies one must to search at the angular distance $18° \geq \beta > 0°$ from the GRB arrival direction. The discovery of the host galaxy by the GRB arrival direction is the artifact.

Observatory (CGRO). The more accurate measurement

The present paper considers the GRB clusters from the Galactic BH binaries GRO J0422+32, XTE J1118+480, Cyg X-1 and others. In this case the coordinates of GRB source are known accurate to one angular second and the reliability of GRB clusters detected from BHBs is without doubts. In the researches of GRB clusters the orderliness in position of successive-in-time GRBs in clusters is found. For each cluster the common typical features, which are related to angular displacement of events, are singled out. The comparison showing the bond of the emissions from the external accretion disk in X-rays and optical range with GRB radiation is fulfilled during the GRO J0422+32 transition into the active state. This comparison allows us to understand the scheme of GRB radiation in general outline.

The paper basis consists of a successive chain of conclusions, in which each following conclusion confirms a previous one: 1) The supposition that the BHB represents a GRB radiating object, was fully confirmed in studying GRBs clusters observed from the Galactic (and extra-Galactic) BHBs; 2) The formation of GRB clusters, associated with photons motion near the compact object, is confirmed by the theory of gravitation, which states that, in such a case, the gravitational dragging effect must be observed. The orderliness of GRBs in clusters is a consequence of the dragging effect confirming that GRBs are radiated near the BH; 3) The dragging effect influence on the orbital motion of photons allows us to imagine the radiation mechanism of bound GRB. The scheme of GRB radiation by the black hole is determined by the dragging force − the Coriolis-type force; 4) The observed dependence of orbital time on the BH mass results in developing the technique of direct measurement of a black hole mass in the BHBs, thereby confirming the previous conclusions; 5) The basic results and their consequences are discussed in the Section 7.

## 2. Identification of GRB Sources with BH Binaries

The basic characteristics of GRB clusters and their possible sources have been discussed till now. In detailed study it was found that GRB clusters from the same source can be observed many times, and the position of GRBs in clusters is ordered. This orderliness consists in the angular displacement of each successive-in-time GRB with respect to preceding one around the source position in the same direction. Just the clusters of bound events, observed, first of all, from Galactic BHBs and presented in Figs. 1, 3, 4, and 5 are considered below.

### 2.1. Black Hole Binary GRO J0422+32

GRO J0422+32 (l", b" = 165.9°, − 11.9°) is a close binary system, in which the BH mass is estimated as 3.6 ± 0.3 $M_\odot$, the mass of an optical star is 0.4 - 0.7 $M_\odot$, the orbital period is 0.2 day, the orbit inclination i = 48°[19].

Table 1 presents the full list of GRBs from the BATSE catalogue [15] for the time 921117 - 940823 that are identified with GRO J0422+32, including the two clusters 1-2-3-4 and *a-b-c-d* (the digit after a point means a serial number of GRB in time for a given day).

**Table 1. GRBs observed from BHB GRO J0422+32**

| # | TJD | GRB date | $t$, hr | $\varphi$, deg | $T_{90}$, s | $\beta$, deg |
|---|---|---|---|---|---|---|
| 1 | 8943 | 921117 | 3.40 | 236 | ~ 0.5 | 14.8 |
| 2 | 8991 | 930104.1 | 1.00 | 67 | 0.6 | 17.2 |
| 3 | 8996 | 930109 | 10.49 | 16 | ~ 40.0 | 12.2 |
| 4 | 9010 | 930123 | 18.76 | 171 | 22.3 | 6.8 |
|   | 9101 | 930424 |  |  | 19.3 | 12.5 |
|   | 9176 | 930708.3 |  |  | 197.0 | 14.6 |
|   | 9193 | 930725.3 |  |  | 0.05 | 14.2 |
| a | 9256 | 930926 | 16.90 | 336 | 89.4 | 2.5 |
| b | 9276 | 931016.2 | 12.55 | 94 | 56.0 | 12.0 |
| b1 | 9279 | 931019.1 | 9.76 | 295 | ? | 17.7 |
| c | 9281 | 931021.2 | 13.01 | 46 | 0.3 | 17.5 |
| d | 9306 | 931115 | 16.56 | 309 | 31.3 | 7.7 |
|   | 9332 | 931211 | 16.15 | 26 | 0.2 | 6.1 |
|   | 9350 | 931229.2 | 16.74 | 190 | ~ 28.0 | 9.5 |
|   | 9405 | 940222 |  |  | 11.1 | 6.3 |
|   | 9442 | 940331.2 |  |  | 28.4 | 15.3 |
|   | 9469 | 940427 |  |  | 4.9 | 11.9 |
|   | 9524 | 940621.2 |  |  | 0.05 | 12.9 |
|   | 9541 | 940708.2 |  |  | 0.2 | 10.4 |
|   | 9587 | 940823.3 |  |  | 19.6 | 6.1 |

Table contains the following information on the GRBs: the time of recording (TJD, date and time *t*), angular position of GRB arrival direction relative to the horizontal axis passing through the source $\varphi$, the GRB duration $T_{90}$, the angular distance $\beta$ between the positions source and GRB. The values $\varphi$ and $\beta$ are calculated by the author. Truncated Julian Day - TJD = JD − 2440000.

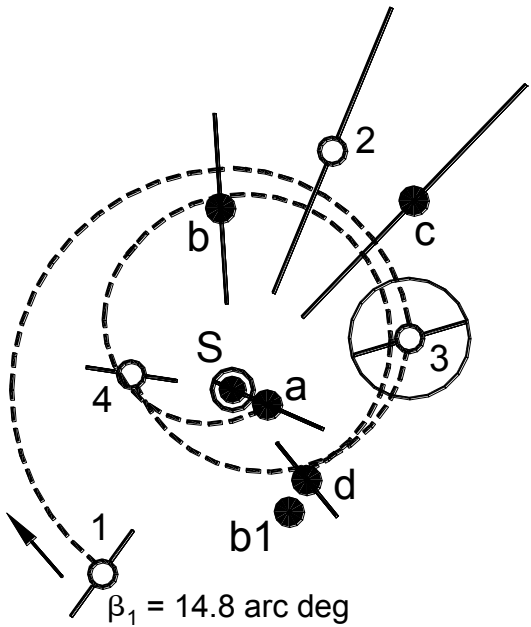

**Fig. 1**. Two GRB clusters observed from GRO J0422+32. The angular distance $\beta_1$ equal 14.8° between the GRB arrival direction and the source position shows the image scale. The position of GRB source is designated by circle with letter S. Error boxes of GRB location are shown by the diameter instead of circle that is explained for GRB # 3.

The disposition of these GRB clusters on the sky sphere is presented in Fig. 1. Each of clusters contains 4 bound bursts 1-2-3-4 and *a-b-c-d*, which are connected by a dashed line. The events follow directly one after another and show angular displacement in the same direction. In all figures the GRB arrival directions are presented, as a rule, in the galactic coordinates taken, from the BATSE catalogue [15].

As a result of angular size conservation the pictures of clusters in Figs. 1, 3, 4, and 5 are similar to that really observed on the sky sphere. These figures are constructed on the basis of the AutoCAD system. From Table 1 mark in GRB clusters from GRO J0422+32 the minimum time between the events and the corresponding angular displacement: between the events 2 and 3 $\Delta t_{min} = t_3 - t_2 = 5.40$ days, $\Delta \varphi = \varphi_3 - \varphi_2 = -51°$; between the events *b* and *c*: $\Delta t_{min} = 5.02$ days, $\Delta \varphi = -48°$.

### 2.2. Observation of Radiations from GRO J0422 + 32 during Transition into Active State

The emission of GRB clusters from GRO J0422+32 coincides in time with the transition of this BHB into the active state. This coincidence allows us to compare in time the radiation of GRB clusters with the radiation observed in the X-ray, optical and radio ranges from the external accretion disk of this BHB.

The binary system BHXN GRO J0422+32 was detected on August 5, 1992 by the BATSE instrument in the ейnergy interval of 20 - 300 keV [20]. The X-ray binaries are mainly at the quiescent state, which lasts from some years to decades. The rate of mass accretion to a primary star at this time is supposed to be below some critical threshold [21, 22], at which the electromagnetic radiation is not observed. The transition into the active state is caused by sharp increase of accretion rate and it is detected by the X-ray outbursts of duration from some weeks to months, which are accompanied by radiation in the other energy ranges. In the outburst onset the luminosity in the X-ray range can increase several orders of magnitude during some days, reaching the value of $10^{37}$ - $5 \times 10^{38}$ erg s$^{-1}$ [23].

Optical radiation from GRO J0422+32 in the active state has been observed during 500 days, the radiation in the radio vawes of 1.5 - 14.9 GHz and in the ultra-violet range - for more than one year [24]. The GRO J0422+32 coordinates were determined from radio emission: $\alpha_{2000} = 4^h 21^m 42^s.75$, $\delta_{2000} = +32°54'27.2"$ with error of 0.5". Optical measurement in the R-band and the BATSE data are presented [25], UBV observations – [26]. The CGRO Compton telescope (COMPTEL) during the time interval from August 12, 1992 to August 20, 1992 has recorded gamma radiation up to 2 MeV [27]. All these long outbursts in the energy range from radio ($10^{-5}$ eV) up to gamma emission ($10^6$ eV) are radiated from the accretion disk, as it is proposed [19].

Figure 2 shows the profiles of light curves in the X-ray range of 20-100 keV and in the optical R - band [19]. The light curve in the X-ray range of 20 - 300 keV, since the detection moment on August 5, 1992, reaches its major maximum of ~ 3 Crab on August 8, 1992. The total duration of the X-ray burst is ~ 200 days. In addition to the primary optical burst, at least two mini-outbursts were observed. According to BATSE catalogue Fig. 2 shows all GRBs, which are radiated from the area concluded in the circle with R = 18° around the position of GRO J0422+32 during nearly 650 days from 921117 to 940823. On the average, the rate of GRB radiation during the time since October 1992 to May 2000 equals one event per 50 days. It is of interest that before transition into the active state of the given binary system, according to BATSE catalogue, the GRBs have not been observed from this area for almost 300 days - since 920122 up to 921110 (TJD 8643 - 8936). One can notice also the absence of GRBs over the X-ray outburst maximum, but, indeed, during this time the events have been observed in a huge amount. The emission of events at a maximum rate, when only within August 6-9, 1992 (TJD 8841 - 8844) the BATSE instrument has recorded 162 events from the area with radius of 20° around the GRO J0422+32 position, exactly coincides in time with the X-ray burst maximum. In total, during 60 days (TJD 8840 – 8900) about 400 weak in the flux events, similar to weakest GRBs, were observed [28]. All these bursts were excluded from the BATSE catalogue by reason of that their origin was associated with GRO J0422+32? The results of this paper, establishing the characteristic features of GRBs and their source, compel us to confirm that indicated 400 events are the GRBs and (or) their counterparts. This statement can be supplemented by the supposition [29] that the Cyg X-1 radiates GRBs observed from the area with angular size of 30° near Cyg X-1. The authors show that some weak events, radiated from the area of Cyg X-1 and excluded from the BATSE catalogue, represent the "classical" GRBs.

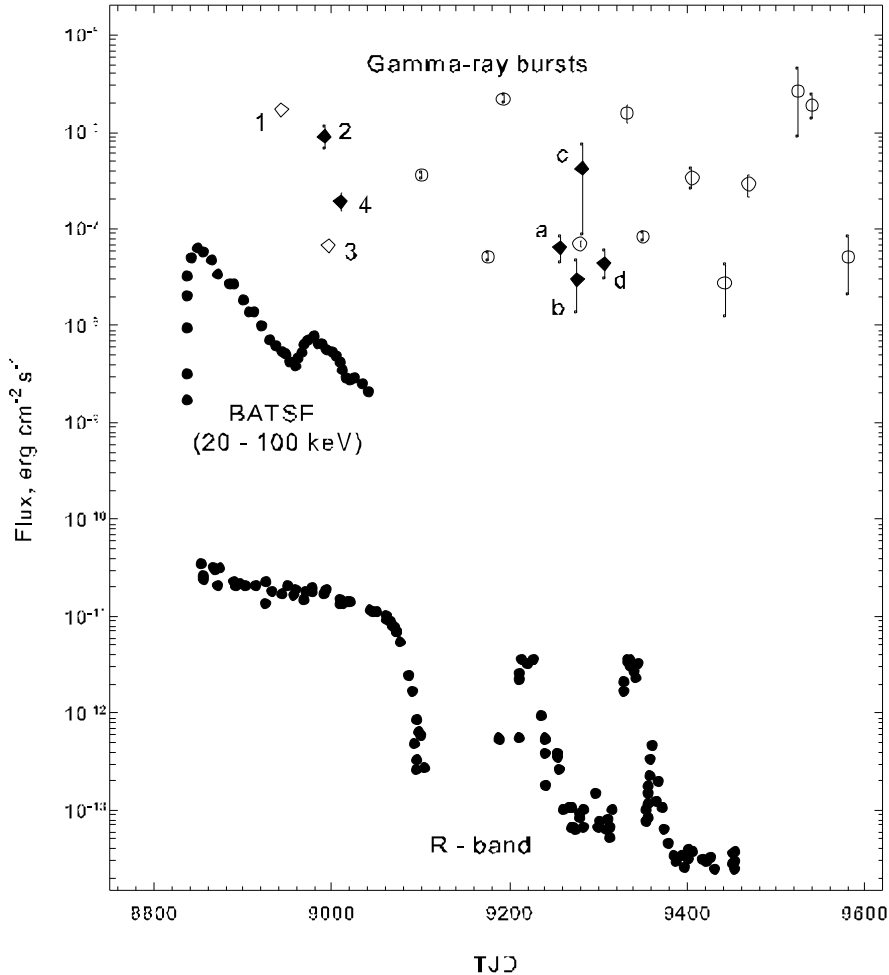

**Fig. 2.** The picture of outbursts in X-ray and optical ranges from GRO J0422+32 is supplemented by GRBs from Table 1. GRBs in clusters are designated by diamonds, the other GRBs - by empty circles; the empty diamonds 1 and 3 present GRBs with the approximate estimate of a flux.

Obviously, the generation of GRB clusters depends on the process of substance accretion, which can have an avalanche character, in particular, during the BHB transition into the active state. Rare GRBs observed from BHBs, apparently, indicate an episodic, weaker accretion during the quiescent state of BHBs. The processes resulting in the disk instability are discussed in reviews [19, 30].

The conclusion, which follows from consideration of processes occurring in BHBs, consists in that the primary electromagnetic radiation from the accretion disk is captured by BH, and we observe these photons at the BH orbits, from which they are re-emitted. The characteristics of secondary radiation (the photon energy and duration of emission) are changed as affected of the BH gravitational potential and the gravitational time slowdown. It is important to note that the GRBs represent the physical process, which constantly occurs near BH in the strong gravitational field.

**2.3. Observation of GRB Clusters from BH Binaries**

We shall consider the GRB clusters observed, first of all, from the Galactic BHBs XTE J1118+480, Cyg X-1, etc., which show the specific features noticed in clusters observed from the binary system GRO J0422+32. As a rule, one can single out two or more clusters from each source.

**2.3.1. Black Hole Binary XTE J1118+480**

Fig. 3 shows the disposition on the sky sphere the GRB cluster from XTE J1118+480 (l", b" = 157.4°, 62.6°), which is of interest as the first BHB found in Galactic halo. According to [31] the characteristics of this binary system are as follows: the BH mass is 6.0 - 7.7 $M_\odot$, the optical star mass is 0.1 - 0.5 $M_\odot$, the orbital period equals 0.17 day, i = 81°.

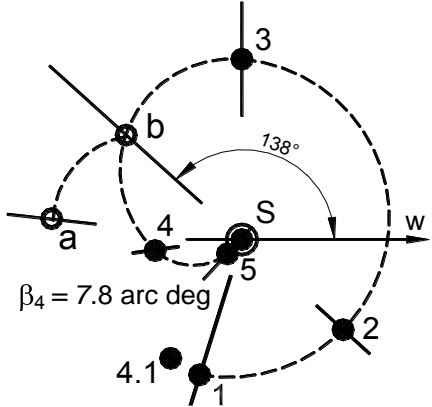

**Fig. 3**. GRB cluster observed from XTE J1118+480. The count of angle $\varphi$ for GRB arrival direction relative to some axis $w$ is shown.

The estimation of the distance to XTE J1118+480 gives D = 1.9 kpc corresponding to a height of 1.7 kpc above the Galactic plane. The characteristics of GRB cluster observed from J1118+480 are given in Table 2.

In accordance with Table 2 $\Delta t_{min}$ and $\Delta\varphi$ between the events 4 and 5 equal 5.5 days, 38°; $\Delta t_{min}$ and $\Delta\varphi$ between the events $a$ and $b$ equal 9.7 days, 36°.

**Table 2. GRB clusters observed from XTE J1118+480**

| # | TJD | GRB | $t$, hr | $\varphi$, deg | $\beta$, deg |
|---|---|---|---|---|---|
| 1 | 9935 | 950806.2 | 20.15 | 254 | 12.8 |
| 2 | 9946 | 950817 | 20.32 | 316 | 12.2 |
| 3 | 9964 | 950904.1 | 7.72 | 90 | 16.4 |
| 4 | 9981 | 950921.1 | 5.82 | 187 | 7.8 |
| 4.1 | 9982 | 950922.2 | 7.20 | 241 | 12.6 |
| 5 | 9986 | 950926 | 17.76 | 225 | 1.9 |
| a | 10333 | 960907.1 | 6.84 | 174 | 16.9 |
| b | 10342 | 960916 | 23.81 | 138 | 13.4 |

**2.3.2. Black Hole Binary Cyg X-1**

The massive binary system Cyg X-1 (l", b" = 71.3°, 3.1°), in which the BH mass is estimated as 10.1 $M_\odot$, the optical star mass is 17.8 $M_\odot$, the orbital period is 5 days, i = 35°, the distance - 2.1 kpc [32]. The Cyg X-1 is the permanent source of X-ray radiation and one of brightest sources of soft gamma-radiation (> 25 keV) possessing high variability. The characteristics of GRBs observed from Cyg X-1 in 2 clusters are given in Table 3. Fig. 4 presents the disposition of the GRB clusters on the sky sphere.

**Table 3. Two GRB clusters observed from Cyg X-1**

| # | TJD | GRB date | $t$, hr | $\varphi$, deg | $\beta$, deg |
|---|---|---|---|---|---|
| a | 8584 | 911124 | 10.70 | 314 | 16.8 |
| a1 | 8587 | 911127.2 | 7.02 | 253 | 13.7 |
| b | 8597 | 911207 | 9.85 | 261 | 8.8 |
| 1 | 9493 | 940521.2 | 15.76 | 41 | 15.7 |
| 2 | 9509 | 940606 | 12.44 | 359 | 14.8 |
| 3 | 9541 | 940708.3 | 20.70 | 219 | 10.7 |
| 4 | 9572 | 940808 | 14.82 | 66 | 12.9 |

From Table 3 the minimum times and corresponding angular displacements between the events $a$ and $b$, 1 and 2 are as follows: $\Delta t$ = 12.96 days, $\Delta\varphi$ = − 53°; $\Delta t$ = 15.86 days, $\Delta\varphi$ = − 42°.

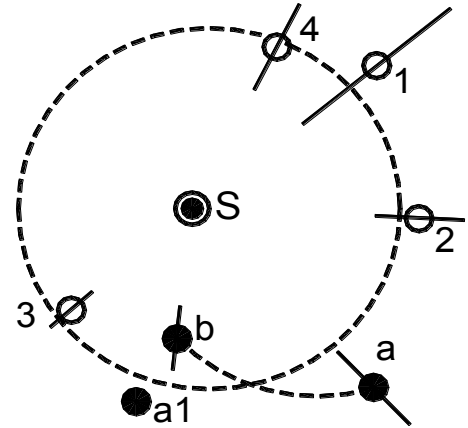

**Fig. 4**. Two GRB clusters observed from BHB Cyg X-1.

### 2.3.3. Unknown BH Binaries

From the catalogue of GRB clusters [14] one can distinguish the unusual GRB clusters from two earlier unknown, possibly, extra-Galactic sources (BHBs). Their coordinates were determined with error of ~ 5°. For the first source with coordinates l", b" = 235°, 49° the minimum times between the bound events in two clusters are 112.1 and 112.6 days and the corresponding angular displacements equal 52° and 45°. The characteristics of GRBs in two clusters observed from second source with coordinates l", b"= 352°, − 32° are given in Table 4, and the cluster disposition on the sky sphere is shown in Fig. 5. The minimum time interval and corresponding angular displacement equal between the events No. 1 and 2: $\Delta t$ = 11.574 days, $\Delta \varphi$ = 42°, and between the events No. 4 and 5: $\Delta t$ = 11.569 days, $\Delta \varphi$ = 47°. These "twins − clusters" allow us to suppose that the GRB clusters of the each source have similar disposition of events (the GRB arrival directions) relative to the source position.

**Table 4. Two GRB clusters observed from BHB (l", b" = 352°, − 32°)**

| # | TJD | GRB | $t$, hr | $\varphi$, deg | $\beta$, deg |
|---|---|---|---|---|---|
| 1 | 8706 | 920325.1 | 12.97 | 112 | 5.0 |
| 2 | 8718 | 920406 | 2.74 | 154 | 12.2 |
| 2.1 | 8726 | 920414.1 | 18.31 | 180 | 17.6 |
| 3 | 8740 | 920428 | 11.30 | 3 | 11.7 |
| 4 | 8767 | 920525.2 | 3.45 | 21 | 5.6 |
| 5 | 8778 | 920605 | 17.10 | 68 | 10.0 |
| 5.1 | 8786 | 920613.1 | 14.53 | 219 | 10.9 |
| 6 | 8791 | 920618 | 21.84 | 281 | 10.3 |
| 7 | 8811 | 920708 | 13.39 | 247 | 5.4 |

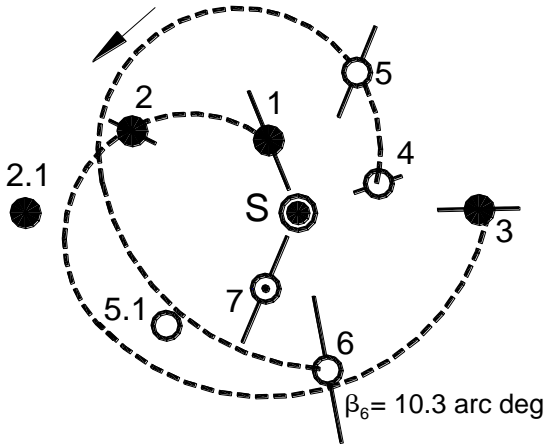

**Fig. 5.** Two GRB clusters observed from BHB with the coordinates l", b" = 352°, −32°. Each of these clusters consists of 3 bound events, both clusters have the similar position relative to the source one, so they are combined each with other in turning one of them on the angle 90° relative to the source location. The identity of these clusters is also confirmed by that the minimum times $\Delta t$ in them between events 1 and 2, 4 and 5 differ by 7 minutes only.

In considered 9 clusters from 5 BHBs the angular displacement $\Delta \varphi$ corresponding to the minimum $\Delta t$ between the bound events does not virtually change. The mean value of $\Delta \varphi$ calculated for 9 GRB clusters equals 46.4° in magnitude. Probably, this angle indicates at the special conditions of GRB generation. The minimum $\Delta t$ between the bound events in clusters, provided that the angular displacement is ~ 46°, varies from 5 to 112 days, i.e. $\Delta t$ has its characteristic value for each BH.

The minimum $\Delta t$ between the bound events establishes also for each BH the time bound defining the doublets of events. The detection of two successive-in-time events divided by the time shorter than minimum $\Delta t$ implies that the second event does not submit the above mentioned orderliness: it has arbitrary position in a cluster. Such pairs (trio) are defined as the doublets (triplet) of events. Examples of such doublets are GRBs 931016.2(*b*) and 931019.1(*b*1) in Table 1 and Fig. 1, as well as the events 2 and 2.1, 5 and 5.1 in Table 4 and Fig. 5. These nonbound GRBs can represent, in principle, the background events, i.e. the events from other sources.

Obviously, that the GRB counterparts also form the clusters almost simultaneously with the GRB clusters.

The discovery in Galaxy the BHBs radiating GRBs confirms the supposition spoken out in Introduction about the erroneous interpretation of the observational facts. The fact that the objects radiating the cosmic GRBs are the BHBs, including the Galactic BHBs, excludes the possibility of the cosmological GRB origin. So, the interpretation of the observed minimum $z$ ~ 1 in favor of the cosmological GRB origin is incorrect. The possible explanation of the apparent contradiction is given in Summary and conclusions.

### 3. Theoretical Base of GRB Cluster Formation

The found orderliness of GRBs in clusters shows physical bond between these events. Such GRB clusters are of great interest because their properties allow us to study physical characteristics of GRB sources. The observed picture of GRB radiation well conforms with theoretical one presented in the review devoted to motion of particles and photons near the compact objects [33].

In particular, it is noted: "... for constructing astrophysical models directly bound with observations of special interest is the geodesic motion outside the event horizon ... (for example), the propagation of photons, which are emitted in the BH vicinity and, then, either escape to a distant observer, or are captured by a black hole ..., as well as the motion over finite orbits, including the influence of a dragging effect on the orbital motion... In the case of bound orbits the influence of a dragging effect or spin-orbital interaction results in shift the periastrons of Keplerian orbits, and for spherical (r - constant) orbits this influence results in displacement of the nodes (points, in which the orbits are intersect with the equatorial plane of BH) in the direction of BH rotation". In accordance with the theory, GRBs in clusters show bound orbits with displacement of nodes, at which the GRB radiation is observed. Black hole can be briefly defined as a clot of gravitation. One can believe that on the BH orbit there is some cloud of photons - a clot of radiation. At the BH orbits the clot of radiation emits the GRBs at nodes resulting to formation of the GRB clusters. The time measured between successive GRBs (between nodes) represents the time (terrestrial), during which the radiation clot makes the complete revolution around the given BH. Then, the observed orbits should be close to spherical ones, and the dragging effect for the minimum orbit should have the same value close to 46°. It is clear that the minimum-in-time orbits, different for each BH, are determined by its size, which linearly depends on mass.

In the practical GRB model the electromagnetic radiation from the accretion disk is captured into BH orbit, which under the influence of a dragging effect is observed as a few bound quasi-spherical orbits. In the general case, the distant observer finds at the nodes the glow throughout the spectrum of electromagnetic radiation.

## 4. Scheme of Gamma-Ray Burst Radiation

The transition from the Newton law of universal gravitation to the Einstein theory of general relativity (GR) is the transition from the gravistatics to the gravidynamics. In this case the additional gravitational field arises around a massive rotating body along with the usual static field. This additional field reveals itself in that all observers moving relative to a gravitating rotating body undergo effect of the additional force, which depends on the azimuth $\Theta$ determining observer's position relative to the central body, as well as on the magnitude and direction of observer's velocity. According to GR, any rotating gravitating body involves into rotation the surrounding space-time. This dragging effect (the Lense-Thirring effect) belongs to gravidynamical GR predictions. The Kerr metric shows the existence of two physically distinguished surfaces around BH: the static limit surface determined by the equation $r_s = m\,[1 + (1 - a^2 \cos^2\Theta)^{1/2}]$ and the event horizon surface, whose radius is $r_h = m\,[1 + (1 - a^2)^{1/2}]$, where $m$ is the mass of gravitating body and $a$ - the angular momentum per unit mass. All objects behind the static limit surface are obliged to rotate in the direction coinciding with BH rotation. The dragging mechanism principles, which are required for understanding the BH's gamma-ray bursts generation process, are presented in accordance with [34, 35]. The specific component of radial acceleration, caused by source rotation, has in some approximation the form [34]:

$$\left(\frac{d^2 r}{dt^2}\right)_{rot} = \frac{2Gma}{c^2 r^2}(\sin^2\Theta)w \qquad (1)$$

Here $w$ is the instantaneous angular velocity of observer's motion relative to the axis Z ($d\varphi/dt$), G is the gravitational constant.

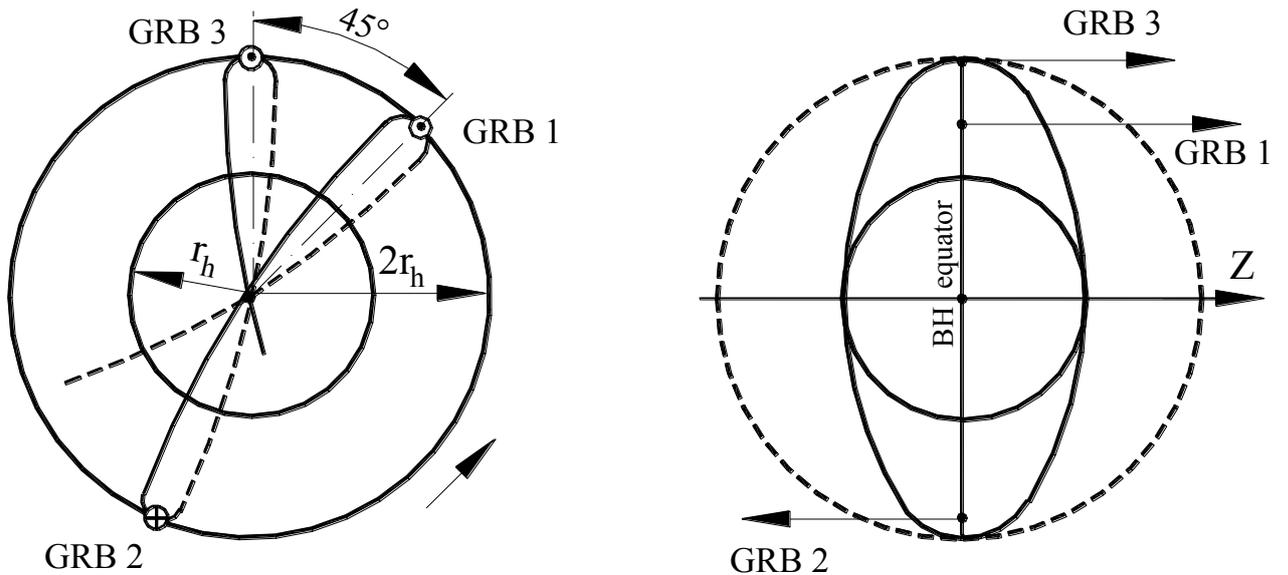

**Fig. 6**. Scheme of GRB radiation by Kerr's black holes. Z is the axis of BH rotation. Photon's bound orbits are designated by solid lines on the visible side and by dashed lines - on the opposite side. The great circle in the left part of the figure and the ellipse on the projection on the right side represent the surface of static limit, and the internal circle represents the event horizon. The dashed circle in the right part of a figure shows the Schwarzschild sphere on whose surface the indicated orbits are located. GRBs are emitted at nodes, i.e. twice per 1/2 orbit: first time - in the observer direction GRB 1, and second time - in the opposite direction GRB 2, which an observer cannot see.

The idealized scheme of GRB radiation by a black hole, which follows from observations, is presented in Fig. 6. The drawing is constructed for the dragging effect resulting in the node displacement by 45° per one orbit, when the GRB 1 and GRB 3, seen by an observer, are radiated. As well, from Fig.6 we can see that the node displacement occurs simultaneously with the orbit turn in accordance with the change of dragging effect in each point of the orbit. Generally, the bound orbits must be quasi-spherical and pass near the Schwarzschild surface. Probably, the given scheme shows the base of process, which results in formation of the jets observed from BH binaries.

The true directions of GRB radiations are differed from indicated ones by arrows in Fig. 6. Indeed, these directions are determined by the orbit radius. The radiation angle of the observed GRBs relative to BH axis rotation is in the limits of $18° \geq \beta > 0°$. The GRBs with other angles are not observed, probably, they are absorbed by BH or gave out from the observer field of view. The minimum angle of GRB radiation ~ 0° corresponds to the orbit the most distant from BH. It is that rare case, when the GRB arrival direction one can coincide with the host galaxy within the accuracy of observation, for example, in Figs. 1 and 3 (tables 1 and 2) this accuracy is 2° and 2.5°, respectively. The $\beta$ angle, as it is accepted in the paper, equals to angular distance (indicated in Tables 1, 2, 3, and 4) between the source location and GRB arrival direction.

## 5. Practical Estimation of Dragging Effect

One can suppose from the theoretical concept of the GRB phenomenon that the polar minimum orbit should be spherical and located as close to the BH as possible. All further estimations were made for extremely rotating BHs (the specific angular momentum $a = 1$), for which the measurement of the event horizon radius is simultaneously the determination of mass $r_h = m$.

The radius of BH horizon is $r_h = m$, and the maximum size of the static limit $r_s = 2m = 2r_h$ is obtained from the formula: $r_s = m [1 + (1 - a^2 \cos^2 \Theta)^{1/2}]$, taking into account $\Theta = 90°$. For radiation moving over the spherical orbit with radius $r_{grb} = 2r_h$ the node - the point of GRB emission - lies at the intersection of the static limit surface with the equatorial plane of BH. Near BH the dragging effect depends on the distance as $r^{-3}$ [36]. The angular velocity of dragging on the event horizon $\Omega_{drag}(r_h)$ equals 360° for one revolution of BH. For $r = r_{grb} = 2r_h$ the angular displacement of the photon orbit equal 45° should be observed for one orbit: $\Omega_{drag}(2r_h) = 360° \times 2^{-3} = 45°$. In other words, during one orbit the node should be displaced exactly by 45° that is confirmed by observations within the accuracy of observation.

The coincidence between theoretical and measured values of the dragging effect leads us to the conclusion that for a minimum orbit the point of GRB emission indeed lies on the static limit surface (see Fig. 6), and its distance from the BH center is $r_{grb} = 2r_h$. If take into account the gravitational radius value $r_g = 2m = 2r_h$, then the minimum spherical orbits will be located on the Schwarzschild surface. This conclusion is, obviously, valid for all extreme BHs independent of their mass, and, accordingly, all BHs in the binary systems radiating GRB clusters represent the BHs with maximum rotation.

Note, that according to the theory for GRBs radiated from $2r_g$ and $3r_g$ the dragging effect must be equal, respectively, 5.6° and 1.7°.

## 6. Direct Measurement of Black Hole Mass

One of practical applications of the observed dragging effect on the orbital motion of the photons is the technique of BH mass measurement. The minimum orbital time, which depends on the BH mass, corresponding to the node displacement equal 45° define two joint criteria, which provide the direct measurement of BH mass. These criteria are determined from the observations and are confirmed by the theory. However, if the orbit measurement error is about a millisecond, then the measurement of a node displacement can have error more than 10° because of rough measurement of the GRB arrival direction. The importance of the 45° criterion consists in that it confirms the validity of a measured minimum orbit.

In the ideal case the time of minimum orbit, disposed on the Schwarzschild sphere, corresponds exactly to the BH mass To measure the BH mass by the proposed technique it is necessary to have the standards for comparison: the mass of BH – $M_{stand}$ and the corresponding time of minimum orbit - $T_{stand}$, the values of which must be known to best accuracy. However, the measured value of BH mass, which could serve as a reference, is absent to date. In calculations the $M_{stand} = 3.5\ M_\odot$ is accepted as the standard mass. Based on observations of GRB clusters the reference minimum time of orbit corresponding to the observed minimum BH mass, which can be accepted for practical calculations, is $T_{stand} = 120$ hours with the error of about minus 1 hour. The standard mass, equal 3.5 $M_\odot$, accepted proceeding from the coincidence of the mass values measured by the given and classical techniques of the BHB GRO J0432+32: $M_{bh} = 3.51 \pm {}^{0}_{0.35}\ M_\odot$, $M_{bh} = 3.57 \pm 0.34\ M_\odot$ [37], and $M_{bh} = 3.66 - 4.97\ M_\odot$ [38]. When discovered of the more reliable $M_{stand}$ and $T_{stand}$ all measured masses one must multiply by some coefficient. The mean error of measurement in the technique is accepted equal 10 % of measured mass on the base of analyze of the few measuring of mass for the same source. When measured of the orbital time ($\Delta$t) the minimum time corresponds to ideal case, when the orbit radius is exactly equal to $r_g$. Practically, in all cases $\Delta$t will be a little more and, respectively, a measured mass will be more than real. This results in to the one-sided error, which is shown further for the measured masses. The accepted value of $M_{stand}$ equal 3.5 $M_\odot$, as show the measurement of BHs in BHBs, can be near to minimum mass of the BH existing at present According to the theory the minimum BH mass must be ≳ 3.0 $M_\odot$. The definition of BH mass is provided by the linear dependence between the mass and measured $T_{min}$: $M_{bh}$ = $kT_{min}$. Here $M_{bh}$ is the black hole mass measured in $M_\odot$; k is the constant coefficient equal to $M_{stand} / T_{stand}$ = 0.7 $M_\odot$ /day; $T_{min}$ is the minimum orbital time for the given black hole measured in days, providing the node displacement is 45°. It should be noted that the reliable measurement of masses this technique provides for BHs with extreme rotation.

Using our technique, the BH masses were calculated for the following binaries:

GRO J0422+32: $M_{bh} = 3.51 \pm^0_{0.35} M_\odot$. This value agrees with $M_{bh} = 3.57 \pm 0.34 M_\odot$ [37] and with $M_{bh} = 3.66 \div 4.97 M_\odot$ [38].

XTE J1118+480: The measured BH mass in this binary system, equal $M_{bh} = 6.8 \pm^0_{0.7} M_\odot$, confirms the value $M_{bh} = 6 \div 7.7 M_\odot$ from calculated [31]. The BH mass in unidentified BHB (p. 3.1) $M_{bh} = 3.9 \pm^0_{0.4} M_\odot$

Cyg X-1: The measured mass of BH $M_{bh} = 9.1 \pm^0_{0.9} M_\odot$ agrees with the value of $6.85 \div 13.25 M_\odot$ [38] and near to the value 10.1 $M_\odot$ [32]. The feature of the BH mass measurement by given technique results in that the measured mass presents the upper limit. For example, the measured BH mass of Syg X-1 equal 9.1 $M_\odot$ defines that the true BH mass is not more than 9.1 $M_\odot$.

GRS 1915+105: The calculated value of the mass $M_{bh}$ = $16.9 \pm^0_{1.7} M_\odot$ corresponds to the value of $M_{bh} = 14 \pm 4$ $M_\odot$ [39, 38].

The masses of BHs for non-identified two BH binaries, which are briefly considered in 2.3.3, were determined. Confirming the uniqueness of the BHB with coordinates l", b" = 352°, − 32°, the definition of BH mass by two clusters gives, practically, the same value: 1) $M_{bh} = 8.1018 \pm^0_{0.81} M_\odot$; 2) $M_{bh} = 8.0983 \pm^0_{0.81} M_\odot$. By definition the minimum of calculated values of BH mass is always taken as a true one. This example shows the possible precision of the BH mass definition by the given technique.

For the second unknown BHB with coordinates l", b" = 235°, 49° the calculated BH mass equals $78 \pm^0_8 M_\odot$. It is of interest in connection with the observed large mass among all known black holes. One can suppose that the optical star also has large mass that can facilitate its observation and, accordingly, identification of the given BHB.

The observed secondary stars in the BHBs are the optical **s**tars of the various spectral classes and masses [19, 30].

Also, the estimations of BH masses for some Galactic BHBs are given in Table 5.

**Table 5**

| BHB | A0620 − 00 | GS 1354 − 64 | GRS 1009 − 45 | GS 2023 + 33 | 4U 1543 − 47 | 4U 1957 + 11 | HMXB 0749 − 00 |
|---|---|---|---|---|---|---|---|
| $M_{bh}, M_\odot$ | $3.47 \pm^0_{0.35}$ | $3.5 \pm^0_{0.4}$ | $4.1 \pm^0_{0.4}$ | $7.1 \pm^0_{0.7}$ | $8.4 \pm^0_{0.8}$ | $8.6 \pm^0_{0.9}$ | $15.7 \pm^0_{1.6}$ |

The weighting of massive black holes ($M_{bh}$ > 100 $M_\odot$) from observation of GRB clusters becomes more complicated in connection with an increasing of observational time that, in its turn, will result in increase of a number of background events, which can mask the real events. The BATSE information on GRB observations during 9 years provides search of the BHs with mass up to ∼ 1000 $M_\odot$. One cannot exclude the probability of GRB clusters radiation by supermassive Kerr's BHs; however, such possibility cannot be easily verified, so for the black hole with mass of $10^4 M_\odot$ the minimum orbit will take 39 years of terrestrial time.

**7. Summary and Conclusions**

The basic problems of astrophysics for recent 35 years were search of black holes and the origin of cosmic GRBs, which were found to be closely related. This paper states that the source of GRB and its counterparts is the BH in binary system and, accordingly, the objects radiating GRBs are the BHBs, including the Galactic BHBs. The electromagnetic radiation observed in the energy range from $10^{-6}$ to ∼ $10^{10}$ eV and, respectively, in the interval of durations from $10^8$ to ≲ $10^{-2}$ s is the electromagnetic glow of BH. The origin of this relativistic phenomenon is all-universe; a BHB at any point of space from Galaxy to the bounds of the observed universe will radiate the GRBs and their counterparts. Then, the observed large-scale isotropy of GRBs (Briggs et al. 1996) is defined by the BHB distribution in the space of universe. The solution of the GRB origin problem was obtained as a result of the discovery of the GRB clusters. Evidently, along with the GRB clusters the similar clusters in the X-ray, optical, radio and other energy ranges exist. The basic characteristics of the observed glow are the energy and duration of radiation Thanks due to these characteristics, GRBs are easy distinguished and detected. In total spectrum of the BH glow the gaps are observed, such that one may to suppose the existence of the glows, in which the GRBs will absent. The BH glow reflects a real picture of process occurring near Kerr's BH with maximum rotation; the picture magnified fantastically large number times, the possible explanation of which is not considered here.

First, it is shown that BHs are the sources of the global electromagnetic radiation. The BH glow occurs in the energy range, exact limits of which unknown at the moment. The correspondence between the energy and the value of gravitation potential in the radiation point gives us a possibility for the theoretical and practical estimations of a maximum value of energy in the BH glow. The photons passing near the BH event horizon must possess of the energy, which is ≳100 GeV from the observations at present.

The paper [40] establishes the dependence between the energy and duration of the BH glow that shows as the energy increases, the duration decreases (and vice versa). As follows from this dependence the GRBs have not any privileged position in observed glow. Also, the dependence of glow duration on the instrument sensitivity is noticed that can transform the short GRBs in the long ones. The results of given paper allow us to imagine in some detail the origin of BH glow. They lead us to the understanding that the photon structure of the BH inner accretion disk has been really shown in the paper [40]. This structure has the clear hierarchy observed on the energy and duration of radiation, which, in its turn, depend on the distance of the radiation point to BH center. Thus, when decreasing of the gravitational potential the energy of BH glow increases and the duration decreases (and vice versa). Then, the duration of BH glow is defined by the gravitational time slowdown. The GRB clusters are radiated from quasi-spherical orbits at r ≥ $r_g$, the spherical orbit at $r_g$ allows us to measure the BH mass. The maximum energy in GRB at $r_g$ is about of 1.5 MeV independent of the BH mass. Thus, at the distance r ≥ $r_g$ the energy of the electromagnetic radiation ($E_{em}$) will be ≤ 1.5 MeV and $E_{em}$ will be > 1.5 MeV at r < $r_g$. In Feynman Lectures on Gravitation [41] is noted: the gravitation forces (potentials) are proportional to value of energy. Thus, the shift of frequency corresponds to part of the gravitational energy in the energy of photon. Evidently, this note relates to influence of the weak field. The BH glow shows that in the field of BH the photon energy, which is observed now up to $5 \times 10^{10}$ eV, completely is defined by the potential and can serve as equivalent of the gravitational energy. It follows from observations that the photons with energy of 2 MeV – 50 GeV are radiated from ergosphere, the glow with energy of ~ $10^{-6}$ eV takes place at the bound of the inner disk. The radiation occurs in the nodes under the action of the dragging force. In general case the wave of radiation is begun in ergosphere and passes throughout the section of the inner disc. The measured at the Earth for 2 GRBs the time delay of the optical radiation (E ~ eV) relative to GRB (E ~ MeV) is about 22 s [42, 43]. Proceeding from the observed dragging effect, one can estimate on the order of magnitude that the bound of inner disk with the hierarchical structure is at the distance of ≲ $10^3$ $r_g$ from the BH (for the minimum BH mass this distance is ≲ $10^4$ km), at which the dragging effect per turn will be $\Omega_{drag}$ ($10^3$ $r_g$) ≲ $3.6 \times 10^{-7}$ degree.

The electromagnetic radiation observed directly from BHB proceeds from the outer accretion disk that allows us to identify of the given BHB as a source of radiation. The BH glow is directly not observed from the BHB and for this reason the GRB source identification with BHBs was delayed almost on 25 years. The identification became possible only after discovery of the GRB clusters. In this case the glow proceeds from the inner disk.

The angular size of GRB clusters and the scheme of GRB radiation allow us to imagine in some approximation the geometry of GRB radiation from the inner disk except an ergosphere. The GRBs are radiated in the opposite directions along the BH rotation axis within the cones with the angle of 36°. The line of sight coinciding with the BH rotation axis defines the best position of observer for detection of GRBs from the given source. In such case the observer detects all GRBs radiated in his semi-sphere and he cannot detect the GRBs radiated in the opposite semi-sphere. Obviously, the GRBs are undetectable from BHBs, when the line observer- source is perpendicular to the BH rotation axis. So, the GRB radiation from the given source can be unavailable for the Earth observer. The angle β between the BH rotation axis and the GRB arrival direction, observed within the limits 18° ≥ β > 0°, defines the limitations on the discovery of GRBs from BHBs. The maximum angle β, which is about of 18°, explains the observed angular size of GRB clusters equal ~ 36°.

Our technique of the BH mass measurement supplements well the classical technique, based on the observations of the characteristics and motion of optical star, and allows us to specify the BH mass. To define the BH mass in binary system we carry out the search of the GRB clusters in the circle with R = 18° around the position of BHB. However, in any domain of such size on the sky sphere about 20 BHBs can be discovered, when limited of the search the BH mass ≤ 100 $M_\odot$. All these sources, practically, coincide by the position and are in the center of circle; then, the total number of sources in this domain will be significantly more. For each of these BHB we can define the BH mass and coordinates accurate to a few degrees. However, for the present we cannot to estimate the distance to these BHBs. The possibility of search of the BHs with any minimum mass is the advantage of our technique. The value 3.5 $M_\odot$ accepted as the minimum mass of BH, resulting from theory and observations can be defined more precisely. The preliminary results, obtained by the given technique, show that the BH masses have the continuous series of values from 3.5 to 1000 $M_\odot$. The indicated maximum value of the measured masses is restricted by the observational time equal 9 years of the BATSE work.

The observation of GRB sources in Galaxy and in the nearest galaxies, for which the redshift is negligible as compared with 1, results in the contradiction with the cosmological interpretation of minimum values z ~ 1 observed from the optical glow. The contradiction compels us to take into account the gravitational redshift, obligatory for given phenomenon, as well as, in accordance with GR, to consider the observed $z$ as the sum of gravitational and cosmological redshifts. Measuring the redshift for the events radiated by Galactic BHBs one can experimentally determine the value of the gravitational redshift, which will be observed in the optical glow of GRB source. In [44] the values of redshift in interval 0.44 – 6.29 measured during the time 970508 – 070810 are given. The all-universe GRB origin (including the Galactic GRB sources) compels us to suppose that the source of GRB 080319B, optical glow of which could be observed with a naked eye, is the near-to-the-Earth Galactic BHB. However, measured for the GRB 080319B redshift equal 0.937 [45] excludes such possibility. According to [44] before observation GRB 080319B two the most bright events with R ≤ 10 mag was known; these are GRBs 990123 and 061007. From them two GRBs 990123 and 080319B have the arrival directions, which are on the angular distance 13°. This value 13° defines the angle between the arrival directions of the indicated GRBs. Knowing that GRBs in clusters from one BHB can be located on the angular distance of 36°, we can assume the possibility of radiation of these GRB by one source. The comparison of GRBs 990123 and 080319B characteristics, as well as of optical glows observed for these events result in the supposition that they are radiated by the same source. The GRB 990123 with duration of 63 s (the BATSE catalogue) is observed in gamma-ray energy above 10 MeV [46]. The GRB coordinates are α, δ = 231.4°, +44.7°, $z$ = 1.6; the glow peak in the visible light corresponds to 8.9 mag [47]. The GRB 080319B with duration of about 60 s is observed in the energy range up to 7 MeV [48]; the GRB coordinates are α, δ = 217.9°, +36.3°, $z$ = 0.937 [49]; the glow peak in the visible light corresponds to 5.3 mag [50]. The basic reason, which excludes the consideration these events as the repeated GRBs of one source, is the difference of their redshifts. To solve this question it is necessary to know the gravitational shifts for these events. The measurement of the gravitational shift (gravshift) for our phenomenon occurring in the field of BH is of principle problem, because all redshifts measured up to now for GRBs (as cosmological ones) demand the correction for the gravitational redshift. The study of the GRB energy spectra shows that the characteristic feature of spectra is their evolution in time. Mainly, the observed variability of spectrum can be the hard-to-soft or soft-to-hard evolution, which, evidently, first was marked [51]. The review about the study of evolution one can find [52]. The results of the given paper allow us to explain of the observed evolution. In the case of hard-to-soft evolution we observe the gravitational redshift and in the case of the soft-to-hard evolution – the gravitational blueshift. It follows that we can measure the gravshift, as there are all necessary observational data for that. As well, remind that the value of gravshift is defined by change of a gravitational potential. The possibility of the gravshift measurement results in the conclusion of interest: if the measured gravitational redshift equals to redshift measured on the optical glow, then this means that the source of given GRB is a Galactic BHB. So, an evolution of spectrum is the very important information as it allows us to estimate the gravitational shift. One may to note, in case the spectrum evolution soft-to-hard is observed one must to search the blueshift on the optical glow.

As it was noted earlier, the identification of the host galaxy location with the GRB arrival direction is the artifact. So, the GRB identification with the true host galaxy remains of the important problem because from the host galaxy we measure the cosmological redshift.

The real BH glow observed, as a rule, in the form of the radiation points, located in the circle with radius equal 18 degrees around the BH, differs essentially from the existing theoretical idea of a glowing aureole around the BH. The radiation mechanism suggests that, depending on the energy, the points of glow will be displaced in coordinates (i. e. the radiation will be observed from some extended area). If the resolution in coordinates of the radiation points can be achieved in the future, then the motion of a radiation point, as a proper motion of a source, can be now observed in the gamma-ray and X-ray ranges, first of all.

The observations of the motion of photon clot at the BH orbit allow us to measure the basic characteristics of extremely fast spinning BHs, for which the knowledge of mass determines automatically their total angular momentum $J = Gm^2 c^{-1}$. As well, the relativistic effects were found, and the BH characteristics predicted by the theory were confirmed, namely: 1) The observation of the black holes with maximum rotation gives an affirmative answer to the question arisen in connection with the BH origin: whether the formation of BH with critical or near-critical angular momentum is natural; 2) The observation of a dragging effect in the strong gravitational field and confirmation of its weakening with a distance near black hole according to the law $r^{-3}$; 3) The demonstration of reality of physically distinguished BH surfaces: the horizon of events, the static limit and Schwarzschild's surface, as well as the possibility of their observation and measurement. These observational facts confirm finally the existence of BHs.

A lot of questions and problems, which follow from the relativistic origin of GRB, demand their explanation and further study: 1) The conservation of GRB clusters angular size independently of a distance to their source; 2) The estimation of the gravitational time slowdown; 3) GRB as an information channel; 4) The possibility of gravitational - wave radiation by BH; 5) The creation of the BH catalogue; and others issues.

The knowledge that GRB source is BH results in one of the most principal consequences of the paper - to the necessity of total refusal from the now dominant doctrine of cosmological GRBs origin. Unfortunately, the concepts of all-universe and cosmological GRB origin are not compatible. Indeed, the BH glow is the relativistic phenomenon, which permanently occurs in the BHB. The reality of the BH glow model is confirmed in that it reasonably explains the all observed features of the phenomenon on the basis of GR. According to the cosmological version, the GRB radiation is the result of a powerful explosive process, a certain Little Bang; the GRBs are radiated at the edge of the observed universe; so that the possibility of GRB origin in Galaxy and in nearby galaxies is excluded. Within the framework of the cosmological concept there are no practical results: the realistic model of the phenomenon is absent; the source of energy, the radiating object, and the GRB radiation mechanism will remain the unknown eternally. Now, when we know the real source of the GRBs and counterparts, the truth is that the almost generally accepted concept of the cosmological GRB origin is a delusion. This, obviously, the greatest delusion in the history of physics will delay the study of the processes occurring in the strong gravitational field and development the contemporary theory of gravitation.

**Acknowledgements** The author thanks V. N. Lukash for the useful discussions on a relativistic side of the problem.